\documentclass[11pt]{article}

\usepackage[a4paper,margin=1in]{geometry}
\usepackage{amsmath,amssymb,amsfonts}
\usepackage{amsthm}
\usepackage{mathrsfs}
\usepackage{bm}
\usepackage{hyperref}
\usepackage{enumerate}
\usepackage{cite}

\theoremstyle{plain}

\theoremstyle{definition}

\theoremstyle{remark}

\begin{document}
	
	\title{\textbf{Finslerian geometrodynamics} }%\\
%		\large Subtitle if Necessary}
	\author{
	Mingwei Zhou\thanks{zhoumw3@mail2.sysu.edu.cn}   \   and\   %\\
	Shi-Dong Liang\thanks{stslsd@mail.sysu.edu.cn}\\	
		\small School of Physics, Sun Yat-Sen University\\
		\small Guangzhou, China
	}
	
	\date{\today}

	\maketitle
	
	% =========================
	% 摘要
	% =========================
	\begin{abstract}
We construct a unified framework of geometrodynamics based on the Finsler geometry to reveal the relationship between spacetime and dynamics.
The Lagrangian of electron in electromagnetic field as the Finsler function gives the
Finslerian metric, which modifies spacetime metric in the Finsler-Randers space. The geodesic equation gives the effective mass, forces, and effective (or geometric) fields. Using the Chern connection, we construct the generalized Einstein-Maxwell equations.
In the local natural basis, we give generalized Maxwell equations and wave equations. We find that the geometric field couples with electromagnetic field and gives effective charges and currents. We analyze several typical cases, such as flat spacetime, vacuum and Berwald structure. We find that the electromagnetic field vanishes, but there still exists the magnetic potential in the Berwald space. These results provide some hints to understand some puzzles, such as axion and dark energy.
These formulations stimulate some clues to explore deeper geometric structures behind physical phenomena.
	\end{abstract}
	
	\tableofcontents
	
	% =========================
	\section{Introduction}
	% =========================
	
The success of Einstein's general relativity reveals the relationship between spacetime geometry and gravity,\cite{Charles}
which stimulates many attempts to generalize the geometric description for understanding some unsolved puzzles, such as the Weyl and Cartan geometries for dark matter and dark energy.\cite{Wu,Xu1,Xu2} Recently, the gravitational spin Hall effect was theoretically predicted based on the geometrodynamics of polarized light.\cite{Oancea,Lars,Pierre} This gives a generalized wave equation coupled with gravity associated with Weyl geometry.\cite{Christos,Hwang,Jacek}

One realized that there exists deep relationships between physics and spacetime geometry.\cite{Girelli,Jacob,Solange,Pfeifer1}
Some novel physical phenomena emerge in deeper geometric structures beyond the Riemannian geometry, such as the Kahler structure and Finsler geometry.\cite{Pettini,Liang2} These geometric frameworks provide a platform to uncover the connections between physical phenomena and mathematical structures.\cite{Bejancu,Miron,Sabau}

In particular, many efforts focus on some novel phenomena in the Finsler geometry, such as the violation of the Lorentz invariance,
\cite{Don,Manuel1,Neil,Schreck,Mignemi,Gibbons1,Kouretsis1,Giovanni,Marco,Iarley} the optimal Zermelo problem,\cite{Gibbons2,Liang1}
the observable effects in the spherically symmetric static Finsler spacetimes and the Finsler-type modification of the Coulomb law.\cite{Claus,Yakov}
Interestingly, the redshift and deflection angle of light was found in the Finsler spacetimes.\cite{Wolfgang,Toshiaki,Elbistan}
The Einstein field equation was generalized to the Finslerian field equation and cosmology.\cite{Kouretsis2,Kouretsis3,Spyros,Li,Li2,Manuel2,Manuel3,Manuel4,Deng,Sjors,Panayiotis,Satoshi1,Satoshi2,Satoshi3}

However, there are still many arguments on how to construct the Einstein field equation because different geometric structures in the Finsler geometry are associated with different connections, such as the Chern, nonlinear Cartan, and Berwald connections.\cite{Andrea1,Andrea2}
In particular, the Finslerin metric defined by the Hessian matrix, in general, does not contain Lorentz signature.
One proposed a triple-Finslerian system $(M,L,F)$ and introduced so-called $d$-tensor to construct the spacetime causal structure with Lorentz signature
for giving the timelike domain in the Finslerian spacetime. \cite{Pfeifer2,Pfeifer3}
Moreover, one found that the Einstein tensor should contain an additional geometric part such that its covariant derivative vanishes for energy-momentum
conservation. \cite{Li2}
Actually, mathematician realize that different definitions of the Ricci curvature tensors turn out different geometric structures of the Finslerian Einstein tensor and field equations. \cite{Esra,Tayebi,Zhiguang}

What physical meanings for these geometric structures of the Finslerian field equation are still unclear.
What relationship between dynamics and spacetime is worth studying further. The interplays between dynamics and geometry can inspire novel insights to dynamical behaviors, such as the Hessian or Jacobi structures,\cite{Bejancu,Miron,Sabau,Shen,Bao}
which provide some hints to explore the interplay between dynamics and geometry for constructing quantum theory of gravity.\cite{Nick,Carlo,Thomas}
In particular, the coupling between gravity and electromagnetic field could reveal the physical scenario of axion on quantum structures in QCD, dark matter and gravitational wave. \cite{Wilczek,Sikivie,Luca1,Luca2}

In this paper, we start from the Finsler-Randers geometry. In the weak field and low velocity approximations, we obtain the Finslerian metric with the Lorentz signature to construct geometrodynamics in the Finsler-Rander geometry, including the Einstein-Maxwell equations and their basic physics.
In Sec. 2, we first review some basic elements on Finsler geometry, including the definition of the Finsler manifold and geodesic equation. In Sec. 3, we start from the Lagrangian of a relativistic electron in electromagnetic field, which can be regarded as the Finsler function in the Finsler-Randers space.
The geodesic equation as the Newton-type equation gives the effective mass, force and relevant fields.
The effective fields involve two geometric structures, dynamics (or Finsler-Randers) and spacetime (or gravity) because the Finsler metric modified spacetime metric.

In Sec. 4, using the Chern connection with the Ricci and scalar curvatures, we construct the Einstein tensor and give the Einstein-Maxwell equation in the Finsler-Randers space. In the local natural basis, we introduce the geometric analogs of the electric and magnetic fields to give the generalized Maxwell and wave equations. The geometric fields couple with electromagnetic fields. The divergence and vorticity of the geometric fields yield effective charges and currents respectively.

In Sec. 5, we discuss a few typical cases, such as flat spacetime, vacuum, and the Berwald structure. Interestingly, in the Berwald space, the electromagnetic field vanishes, but there still exists the magnetic potential. The charge and current drive the effective geometric field. In Sec. 6, we discuss some potential physical applications and perspectives of these formulations. Finally, in Sec. 7,
we give conclusions and outlooks. In the appendix, we give some basic preliminaries on Finsler geometry, including the Chern connection, the Ricci scalar, Ricci curvature, Chern-Riemannian curvature and covariant derivatives. We also give some calculation notes for readers convenience.

	% =========================
	\section{Preliminaries on Finsler geometry}
	% =========================
\subsection{Finsler manifold}\label{subsec2.1}
Let us briefly introduce the Finsler geometry,

Definition: A $n$-dimensional smooth function $F=F(x,y)$, where $y\in T_xM$ and $F:TM\rightarrow [0,\infty )$,
is called the Finsler function (or Finsler metric or structure) if it satisfies the following axioms\cite{Bao}
\begin{itemize}
	\item Positive and smooth: $F(x,y)\geq 0$ and for $y=0$, $F(x,y)=0$ and $F(x,y)\in C^\infty$;
	\item 1-homogeneous: $F(x,\lambda y)=\lambda F(x,y)$, $\forall\lambda\in \mathbb{R}^+$,
	\item Regular: The Hessian matrix defined by
	\begin{equation}\label{FMT1}
		g_{ij}(x,y):=\frac{1}{2}\frac{\partial^2 F^2}{\partial y^i\partial y^j},
	\end{equation}
	is positive-definite for $\forall(x,y)\in TM\backslash 0$.
\end{itemize}
The Hessian matrix $g_{ij}(x,y)$ is called the fundamental tensor (or Finslerian metric) in the Finsler manifold.
A manifold endowed the Finsler metric $(M,F)$ is called the Finsler manifold. Here we use the notation $(x,y)\equiv (x^1,\cdots, x^n, y^1, \cdots, y^n)$ for mathematician convenience. In physics, $y^i=\frac{\partial x^i}{\partial \tau}=v^i$ is velocity.

In particular, when $g_{ij}(x,y)$ is independent of $y$, it reduces to the Riemannian metric $g_{ij}(x)$. The manifold can be regarded as a particular Finsler manifold with a Riemainnian metric (or structure), or the Riemaniann manifold contains a Finsler structure.  In other words,
there exists a relationship between between a Finsler manifold $(M,F)$ and a  Riemannian manifold $(M,g)$.
When the Finsler metric $g_{ij}(x,y)$ is independent of $x$, the metric is called as the Minkowski metric.

It should be remarked that the positive definite Finslerian metric is not necessary for physics. It should be reduced to non-degenerate for describing spacetime geometry, which is called pseudo-Finsler geometry. In the following section, we use actually the pseudo-Finsler geometry.
	
\subsection{Geodesic equation}\label{subsec2.2}
The action of the particle in the nD Finsler manifold is given by
\begin{equation}\label{Act0}
	S=\int F^2(x,y)d\tau.
\end{equation}
Using the variational principle, the Euler-Lagrange equation in the is given by
\begin{equation}\label{ELEq0}
	\frac{d}{d\tau}\frac{\partial F^2}{\partial y^i}-\frac{\partial F^2}{\partial x^i}=0.
\end{equation}
Note that $y^i=\frac{\partial x^i}{\partial \tau}=v^i$ is the velocity in the local coordinate representation.
and
\begin{equation}\label{ELEq1}
	\frac{d}{d\tau}\frac{\partial F^2}{\partial y^i}=\frac{\partial^2 F^2}{\partial y^i\partial y^j}\dot{y}^j
	+\frac{\partial^2 F^2}{\partial y^i\partial x^k}y^k,
\end{equation}
the Euler-Lagrange equation can be rewritten as the geodesic equation,\cite{Bao}
\begin{equation}\label{GDE0}
	\dot{y}^i+2G^i(x,y) = 0,
\end{equation}
where $\dot{y}^i=\frac{dv^i}{d\tau}=\ddot{x}^i$ is the acceleration of the particle and $G^i(x,y)$ is called the geodesic coefficient,\cite{Bao}
\begin{equation}\label{FG3}
	G^i\left(x,y\right)=\frac{1}{4}g^{ij}\left(\frac{\partial^2 F^2}{\partial y^j\partial x^k}y^k-\frac{\partial F^2}{\partial x^j}\right),
\end{equation}
where $(g^{ij})=(g_{ij})^{-1}$. The more details on Finsler geometry are presented in Appendix A or consulting references.\cite{Bao}

	% =========================
	\section{A relativistic electron in Finsler space}
	% =========================
\subsection{Lagrangian, Finsler function and Finslerian metric}\label{subsec3.1}
Let us consider an electron moving in curved space with electromagnetic field, its Lagrangian is given by,\cite{Jackson,Miron,Sabau}
\begin{equation}\label{LG0}
	\mathcal{L}\left(x,\frac{\partial x}{\partial \tau}\right)=mc\sqrt{\widetilde{g}_{\mu\nu}\frac{\partial x^\mu}{\partial \tau}\frac{\partial x^\nu}{\partial \tau}}+\frac{e}{c}A_\mu(x)\frac{\partial x^\mu}{\partial \tau},
\end{equation}
where $x^\mu=(x^0,x^1,x^2,x^3)$ is the 4-position vector and $\tau$ is the proper time.
$A_\mu=(\phi,\mathbf{A})$ is the 4-vector magnetic potential.
The $\widetilde{g}_{\mu\nu}$ is the spacetime metric with the Lorentz signature $(+,-,-,-)$.
The Lagrangian can be regarded as the fundamental function in the Finsler geometry.\cite{Bao,Miron,Sabau}

It should be reminded that we use the physicist notations, $\frac{\partial x^\mu}{\partial \tau}=v^\mu\equiv y^\mu$ from this section. The Lagrangian (\ref{LG0}) plays the role of the Finsler function, $\mathcal{L}:=F$.
Moreover, for convenience, we set $c=1$ and introduce the notations of mathematician,\cite{Bao,Miron,Sabau}
\begin{subequations}\label{LAB}
	\begin{eqnarray}
		\mathcal{L}&=& \alpha+\beta, \\
		\alpha &=&  m\widetilde{\alpha}, \quad \widetilde{\alpha}=\sqrt{\widetilde{g}_{\mu\nu} y^\mu y^\nu},\\
		\beta &=& b_\mu y^\mu, \quad b_\mu\equiv eA_\mu.
	\end{eqnarray}
\end{subequations}
In the weak field approximation, $\|b\|< 1$, which ensures $\mathcal{L}>0$, namely the Lagrangian space $(M,\mathcal{L})$ is referred as to the Finsler-Randers space. \cite{Bao,Miron,Sabau}
The Finslerian metric can be obtained as \cite{Bao}(See Appendix B.1)
\begin{equation}\label{FMT0}
	g_{\mu\nu}=\frac{\mathcal{L}}{2\alpha}\left(m^2\widetilde{g}_{\mu\nu}-\ell_\mu \ell_\nu\right)+(\ell_\mu+b_\mu)(\ell_\nu+b_\nu),
\end{equation}
where $\ell_\mu =\frac{\partial \alpha}{\partial y^\mu}$.

It can be verified that the Finslerian metric is non-degenerate such that the Finslerian metric is invertible.\cite{Bao}
Moreover, the Finslerian metric deforms from the spacetime metric. In the weak field and low velocity approximations, $A_\mu\ll 1$ and $v\ll c$, we assume that the Finslerian metric still remain the Lorentz signature inside the timelike cone of $\widetilde{g}$.\cite{Voicu,Heefer}
The Finslerian metric gives a coupling between dynamics and spacetime or gravity. In other words, the dynamics deforms spacetime in the Finsler-Randers space.

\subsection{Geodesic coefficient}\label{subsec3.2}
The action of this system is defined by $S=\int \mathcal{L}^2(x,y)d\tau$ in Finsler geometry.
By variational calculation, we obtain the Euler-Lagrange equation, which can be rewritten as the geodesic equation,\cite{Bao}
\begin{subequations}\label{GDE1}
	\begin{eqnarray}
		\dot{x}^\mu &=& y^\mu,\\
		\dot{y}^\mu+2G^\mu(x,y) &=& 0,
	\end{eqnarray}
\end{subequations}
where $\dot{x}^\mu=\frac{\partial x^\mu}{\partial \tau}$ and $\dot{y}^\mu=\frac{\partial y^\mu}{\partial \tau}$.
$G^\mu$ is the geodesic coefficient, which can be obtained (See Appendix B.2)

\begin{equation}\label{EMT1}
	G^\mu(x,y)=\frac{1}{4}g^{\mu\nu}\left(-2e\alpha F_{\nu\sigma}+S_{\nu\lambda\sigma}y^\lambda+Q_{\nu\lambda\sigma}y^\lambda
	+M_{\nu\lambda\sigma}y^\lambda\right)y^\sigma,
\end{equation}
where
\begin{subequations}\label{EMF2}
	\begin{eqnarray}
		F_{\nu\sigma}&=& \frac{\partial A_\sigma}{\partial x^\nu}-\frac{\partial A_\nu}{\partial x^\sigma},\\
		S_{\nu\lambda\sigma}&=& 2e^2\frac{\partial \mathcal{A}_{\nu\sigma}}{\partial x^\lambda}-e^2\frac{\partial \mathcal{A}_{\lambda\sigma}}{\partial x^\nu}
		+2e\frac{\partial B_{\lambda\nu}}{\partial x^\sigma},\\
		Q_{\nu\lambda\sigma}&=& m^2\left(\frac{\partial \widetilde{g}_{\nu \lambda}}{\partial x^\sigma}+\frac{\partial \widetilde{g}_{\lambda\nu}}{\partial x^\sigma}
		-\frac{\partial \widetilde{g}_{\lambda\sigma}}{\partial x^\nu} \right),\\
		M_{\nu\lambda\sigma}&=& \frac{em}{\widetilde{\alpha}}\left(A_\nu\frac{\partial}{\partial x^\kappa}-A_\kappa\frac{\partial}{\partial x^\nu}\right)
		\widetilde{g}_{\lambda\sigma}y^\kappa,
	\end{eqnarray}
\end{subequations}
where $F_{\nu\sigma}$ is the electromagnetic field.
$S_{\nu\lambda\sigma}$ contains two kinds of effects, in which $\mathcal{A}_{\nu\sigma}:=A_\nu A_\sigma$ describes the component mixing of the electromagnetic fields and $B_{\lambda\nu}:=A_\lambda \ell_\nu$ describes a coupling of the electromagnetic field and velocity field.
$M_{\nu\lambda\sigma}$ gives a coupled effect between curved spacetime and electromagnetic field.
$Q_{\nu\lambda\sigma}$ depends on curved spacetime.
In the flat spacetime, $Q_{\nu\lambda\sigma}=M_{\nu\lambda\sigma}=0$. When $A_\mu$ vanishes, $F_{\nu\sigma}=M_{\nu\lambda\sigma}=S_{\nu\lambda\sigma}=0$,
but $Q_{\nu\lambda\sigma}\neq 0$.

\subsection{Geometric forces and fields}\label{subsec3.3}
The geodesic equation (\ref{GDE1}b) can be rewritten as the Newton-type equation,
\begin{equation}\label{EGD1}
	m^E_{\mu\nu}\ddot{x}^\nu=f_\mu,
\end{equation}
where $m^E_{\mu\nu}$ is the effective mass,
\begin{equation}\label{Emass1}
	m^E_{\mu\nu}=\frac{2g_{\mu\nu}}{\mathcal{L}}
\end{equation}
and the effective force is given by
\begin{equation}\label{Eforce1}
	f_\mu=-\frac{2g_{\mu\nu}G^\nu}{\mathcal{L}}.
\end{equation}

We may assume that the effective force is generated by an effective field, namely, $f_\mu=\mathcal{F}_{\mu\nu}y^\nu$ as an analog of the electromagnetic field. Thus, the total field can be expressed as the electromagnetic field and the total effective (geometric) field,
\begin{equation}\label{FF1}
	\mathcal{F}_{\mu\nu}=\left(1-\frac{\beta}{\mathcal{L}}\right)F_{\mu\nu}+F^G_{\mu\nu},
\end{equation}
where
\begin{equation}
	F^{G}_{\mu\nu}= -\frac{y^\lambda}{2\mathcal{L}}\left(S_{\mu\lambda\nu}+Q_{\mu\lambda\nu}+M_{\mu\lambda\nu}\right)
\end{equation}
is the total effective (or geometric) field.

\subsection{Geodesic equation beyond Riemannian geometry}
The geodesic coefficient can be expressed as another equivalent form, \cite{Bao}
\begin{equation}\label{GGb0}
	G^\lambda=\frac{1}{2}\gamma^\lambda\ _{\mu\nu} y^\mu y^\nu,
\end{equation}
where
\begin{equation}\label{GamA}
	\gamma^\lambda\ _{\mu\nu}=\frac{1}{2}g^{\lambda \sigma}\left(\frac{\partial g_{\sigma\nu}}{\partial x^\mu }
	+\frac{\partial g_{\mu\sigma}}{\partial x^\nu}-\frac{\partial g_{\mu\nu}}{\partial x^\sigma}\right).
\end{equation}
is the Christoffel symbol of the Finsler geometry. Note that there exists a relationship between two-type Christoffel symbols for the Finsler and Riemannian geometries,\cite{Bao} (See Appendix B.3)
\begin{eqnarray}\label{GGe}
	\gamma^i\ _{jk} y^j y^k &=&\widetilde{\gamma}^i\ _{jk} y^j y^k+m^2B_{j|k}\left(\widetilde{g}^{ij}y^k-\widetilde{g}^{ik}y^j \right)\alpha   \nonumber\\
	&+& B_{j|k}\ell^i \left[y^jy^k+\left(y^j\widetilde{b}^k-y^k \widetilde{b}^j\right)\alpha\right],
\end{eqnarray}
where
\begin{equation}\label{GamA1}
	\widetilde{\gamma}^i\ _{jk} = \frac{1}{2}\widetilde{g}^{i \ell}\left(\frac{\partial \widetilde{g}_{\ell k}}{\partial x^j}
	+\frac{\partial \widetilde{g}_{j\ell}}{\partial x^k}-\frac{\partial \widetilde{g}_{jk}}{\partial x^\ell}\right)
\end{equation}
is the Christoffel symbol of spacetime geometry, namely the spacetime connection coefficient, and
\begin{equation}\label{BbB0}
	B_{j|k}:=\frac{\partial b_j}{\partial x^k}-b_\sigma\widetilde{\gamma}^\sigma\ _{jk}.
\end{equation}
Consequently, the geodesic equation can be expressed as two-type equivalent forms based on two geometries,
\begin{subequations}\label{GDEa}
	\begin{eqnarray}
		\dot{y}^\mu+\gamma^\mu\ _{\rho\sigma} y^\rho y^\sigma &=& 0, \\
		\dot{y}^\mu+\widetilde{\gamma}^\mu\ _{\rho\sigma} y^\rho y^\sigma &=& \mathcal{B}^\mu,
	\end{eqnarray}
\end{subequations}
where
\begin{eqnarray}\label{BBa}
	\mathcal{B}^\mu &:=& -m^2B_{j|k}\left(\widetilde{g}^{\mu j}y^k-\widetilde{g}^{\mu k}y^j \right)\alpha \nonumber\\
	&-& B_{j|k}\ell^\mu \left[y^jy^k +\left(y^j\widetilde{b}^k-y^k \widetilde{b}^j\right)\alpha\right],
\end{eqnarray}
describes the geodesic deviation of the Finsler geometry from the Riemannian geometry.
When $B_{j|k}=0$, the geodesic coefficient can be expressed as
\begin{equation}\label{GGfa}
	G^i = \frac{1}{2}\gamma^\mu\ _{jk} y^j y^k= \frac{1}{2}\widetilde{\gamma}^i\ _{jk} y^j y^k.
\end{equation}
The two-type geodesic equations (\ref{GDEa}) reduce to consistence. The condition $B_{j|k}=0$ is so-called the Berwald space of the Finsler geometry.
The Berwald space is a subspace of the Finsler space beyond the Riemannian geometry.

	% =========================
	\section{Generalized Einstein-Maxwell equations}
	% =========================
	
\subsection{Finslerian spacetime and energy conservation}\label{subsec4.2}
In principle, constructing Einstein field equation in Finsler spacetime should obeys two physical laws and one mathematical fact.
\begin{itemize}
	\item The Finslerian spacetime should hold causality, namely the Finslerian metric should have Lorentzian signature such that physical events happen in the timelike domain.\cite{Pfeifer2,Pfeifer3}
	\item  The covariant divergence of the Einstein tensor should vanish identically for energy-momentum conservation.\cite{Li2}
	\item  In Finsler geometry, the Ricci and scalar curvatures depend on connections, such as Chern, Cartan and Berwald connections.
\end{itemize}
However, the Finslerian metric $g_{\mu\nu}$ in general does not fall in the domain with the Lorentzian signature. One proposed a triple-Finslerian system to introduce so-called $d$-tensor to give the causal structure and the timelike cones as well as observer frame.\cite{Pfeifer2,Pfeifer3}
Here, we adopt a perturbation method to approximately remain the Lorentzian signature of the Finslerian metric. The Finslerian metric (\ref{FMT0}) can be rewritten as
\begin{equation}\label{FMTa}
	g_{\mu\nu}=\frac{1}{2}\left(1+\frac{\beta}{\alpha}\right)m^2\widetilde{g}_{\mu\nu}+\theta_{\mu\nu},
\end{equation}
where $\theta_{\mu\nu}:=\frac{1}{2}\left(1-\frac{\beta}{\alpha}\right)\ell_\mu \ell_\nu+\ell_\mu b_\nu+\ell_\nu b_\mu+b_\mu b_\nu$.
In the weak field and low velocity approximations, $A_\mu\ll 1$ and $y\ll c$, it can be verified numerically that the Finslerian metric remains Lorentzian signature of $\widetilde{g}_{\mu\nu}$ provided $|\theta_{\mu\nu}|\ll \frac{1}{2}\left(1+\frac{\beta}{\alpha}\right)m^2|\widetilde{g}_{\mu\nu}|$.
Moreover, here we adopt the Chern connection (torsion free) to construct the Ricci and scalar curvatures.\cite{Li2, Bao}

On the other hand, the covariant divergence of the conventional Einstein tensor in the pseudo-Riemannian geometry does not vanish. In principle, using the second Bianchi identities to construct the Einstein tensor for given connection.\cite{Li2} To reveal basic features of the Einstein tensor in the Finslerian-Randers space, we consider a particular case called the Berwald space to construct the Einstein tensor,\cite{Li2} in which the $hv$-part of the Chern curvature vanishes, which is equivalent to $B_{j|k}=0$ in (\ref{BbB0}). (see Appendix B.3)

\subsection{Generalized Einsten-Maxwell equations}\label{subsec4.2}
Let us first recall the Einstein-Maxwell equations in the pseudo-Riemannain geometry, \cite{Charles}
\begin{subequations}\label{EMeq0}
	\begin{eqnarray}
		G_{\mu\nu} &=& \kappa T_{\mu\nu}, \\
		\partial_\mu F^{\mu\nu}&=& J^\nu,\\
		\partial_\lambda F^{\mu\nu}+\partial_\mu F^{\nu\lambda}+\partial_\nu F^{\lambda\mu} &=& 0,
	\end{eqnarray}
\end{subequations}
where $\kappa=\frac{8\pi G}{c^4}$.  The Einstein tensor is given by
\begin{equation}\label{ET0}
	G_{\mu\nu}=R_{\mu\nu}-\frac{1}{2}\widetilde{g}_{\mu\nu}R,
\end{equation}
and the energy-momentum tensor is
\begin{equation}\label{EMT0}
	T_{\mu\nu}=F_{\mu\sigma}F^{\sigma}\ _\nu-\frac{1}{4}\widetilde{g}_{\mu\nu}F_{\alpha\beta}F^{\alpha\beta}.
\end{equation}

To generalize the Einstein-Maxwell equation to those in the Finsler-Randers space, we should consider the Finsler-Randers features and physical laws mentioned above. When the spacetime metric is generalized to the Finslerian metric, the Ricci and scalar curvatures depend on what connection we choose, which turns out the connection-dependent Bianchi identities and modifies the Einstein tensor. To reveal some basic physics in the Finsler-Randers space, here we choose the Chern connection with the Berwald structure, namely, $B_{j|k}=0$ in (\ref{BbB0}).\cite{Li2}
Using the second Bianchi identities to remain the energy-momentum conservation, the Einstein tensor can be constructed by\cite{Li2}
\begin{equation}\label{EST0}
	G_{\mu\nu}:=Ric_{\mu\nu}-\frac{1}{2}g_{\mu\nu}S+\Lambda_{\mu\nu},
\end{equation}
where $Ric_{\mu\nu}$ is the Ricci curvature tensor and $S$ is the scalar curvature. $\Lambda_{\mu\nu}$ is introduced for energy-momentum conservation.
The Ricci curvature tensor is defined by\cite{Bao}
\begin{equation}\label{Rcci2}
	Ric_{\mu\nu}:=\frac{1}{2}\frac{\partial^2 Ric}{\partial y^\mu \partial y^\nu}.
\end{equation}
where $Ric$ is the Ricci scalar curvature defined by\cite{Bao}
\begin{equation}\label{RCC1}
	Ric=R^\mu\ _{\mu}=2\frac{\partial G^\mu}{\partial x^\mu}-y^\kappa \frac{\partial^2 G^\mu}{\partial x^\kappa\partial y^\mu}
	+2G^\kappa\frac{\partial^2 G^\mu}{\partial y^\kappa\partial y^\mu}
	-\frac{\partial G^\mu}{\partial y^\kappa}\frac{\partial G^\kappa}{\partial y^\mu}.
\end{equation}
The scalar curvature is defined by $S:=g^{\mu\nu}Ric_{\mu\nu}$, and the additional term is given by \cite{Li2}
\begin{equation}\label{BbB}
	\Lambda_{\mu\nu}:=\frac{1}{2}B_\kappa\ ^\kappa\ _{\mu\nu}+B_\mu\ ^\kappa\ _{\nu\kappa}
\end{equation}
where
\begin{subequations}
	\begin{eqnarray}\label{Bbb}
		B_\kappa\ ^\sigma\ _{\mu\nu} &=& g^{\sigma\lambda}B_{\kappa\lambda\mu\nu}, \quad
		B_{\kappa\lambda\mu\nu}:=-A_{\kappa\lambda\sigma}R^\sigma_{\mu\nu}, \\
		A_{\kappa\lambda\sigma}&:=& \frac{\mathcal{L}}{2}\frac{\partial g_{\kappa\lambda}}{\partial y^\sigma}, \qquad
		R^\sigma_{\mu\nu} := \ell^\kappa R_\kappa\ ^\sigma\ _{\mu\nu},
	\end{eqnarray}
\end{subequations}
where $A_{\kappa\lambda\sigma}$ is the Cartan tensor and $\ell^\kappa:=\frac{y^\kappa}{\mathcal{L}}$.
It can be verified that the covariant divergence of the Einstein tensor is zero, $G_{\mu\nu|\mu}=0$, and $G_{\mu\nu;\mu}=0$,\cite{Li2} which implies energy-momentum conservation, $T_{\mu\nu|\mu}=0$ and $T_{\mu\nu;\mu}=0$.\cite{Li2} The additional tensor $\Lambda_{\mu\nu}$ depends on the Cartan connection coefficient, which can be interpreted as a geometric description of cosmological constant. This geometric feature of cosmological constant could provide a geometric scenario of dark energy.

In order to generalize the Maxwell equations to those in the Finsler-Randers space, suppose that the Lagrangian of the total field is given by
\begin{equation}\label{LFF}
	\mathcal{L}_F=-\frac{\sqrt{-g}}{4}\mathcal{F}_{\mu\nu}\mathcal{F}^{\mu\nu}+\sqrt{-g} A_{\kappa}J^\kappa,
\end{equation}
where we set the coupling constant $c\equiv 1$ and $\mathcal{F}_{\mu\nu}=F_{\mu\nu}+F^G_{\mu\nu}$ is the total field, where we ignore the factor $\frac{\beta}{\mathcal{L}}$ in (\ref{FF1}) in the weak field and low velocity approximations without lose of key physics. Suppose that $\nabla_\mu \sqrt{-g}=0$ and using the Euler-Lagrange equation we can obtain the generalized Maxwell equation,

\begin{equation}\label{GMWEq1}
	\nabla_\mu F^{\mu\nu}+\nabla_\mu \mathcal{F}^{G\left[\mu\nu\right]}
	-\frac{1}{4}\nabla_\mu\left(\frac{\partial F^G_{\alpha\beta}}{\partial(\partial_\nu A_\mu)} \mathcal{F}^{\alpha\beta}+\mathcal{F}_{\alpha\beta}\frac{\partial F^{G\alpha\beta}}{\partial (\partial_\nu A_\mu)}\right) =J^\nu,
\end{equation}
where $\nabla_\mu$ is the covariant derivative in the horizontal direction, which is defined by 
\begin{equation}
		\nabla_\mu \mathcal{F}^{\mu\nu}=\frac{\delta \mathcal{F}^{\mu\nu}}{\delta x^\mu}
		+\mathcal{F}^{\kappa\nu}\Gamma^\mu\ _{\kappa\mu}
		+\mathcal{F}^{\mu\kappa}\Gamma^\nu\ _{\kappa\mu}
\end{equation}
where $\Gamma^\mu\ _{\kappa\mu}$ is the Chern connection coefficient and
\begin{equation}
	\frac{\delta }{\delta x^\mu}=\frac{\partial}{\partial x^\mu}-N^\kappa\ _\mu \frac{\partial}{\partial y^\kappa} 
\end{equation}
where $N^\kappa\ _\mu$ is the Cartan connection coefficient. 

It can be seen that the Finslerian effects modify the Maxwell equations, including the spatial derivative of the geometric field, the couplings between the electromagnetic field and geometric field as well as their derivatives with respect to $A_\mu$ and $\partial_\nu A_\mu$. The source Maxwell equations depend on the Finslerian structure, involving the Chern and Cartan connection coefficients. However, 
the source-free electromagnetic field equation remains the same form because it obeys the Bianchi identity,
\begin{equation}\label{HDF1}
	\frac{\partial F^{\mu\nu}}{\partial x^\lambda}+\frac{\partial F^{\lambda\mu}}{\partial x^\nu}+\frac{\partial F^{\nu\lambda}}{\partial x^\mu} =0.
\end{equation}
In order to get the basic features of the source Maxwell equations with the geometric field, we ignore the detail of the Finslerian structure, $\Gamma^\lambda\ _{\mu\nu}$ and $N^\mu\ _\nu$. Consequently, we establish the Einstein-Maxwell equation, (see Appendix B.4)
\begin{subequations}\label{EMeq2}
	\begin{eqnarray}
		Ric_{\mu\nu}-\frac{1}{2}g_{\mu\nu}S &+& \Lambda_{\mu\nu}=\kappa T_{\mu\nu}, \\
		\frac{\partial F^{\mu\nu}}{\partial x^\nu}+\frac{\partial F^{G[\mu\nu]}}{\partial x^\nu} 	&-&\frac{1}{2} \mathcal{F}^{\rho\sigma}
		\frac{\partial^2 F^G_{\rho\sigma}}{\partial x^\nu(\partial_\nu A_\mu)} \\
		&-&\frac{1}{4}\left( \frac{\partial \mathcal{F}^{\rho\sigma}}{\partial x^\nu}
		 \frac{\partial F^G_{\rho\sigma}}{\partial (\partial_\nu A_\mu)}
		 +\frac{\partial \mathcal{F}_{\rho\sigma}}{\partial x^\nu}
		 \frac{\partial F^{G\rho\sigma}}{\partial (\partial_\nu A_\mu)}\right) 
		=J^\mu, \\
		\frac{\partial F^{\mu\nu}}{\partial x^\lambda}+\frac{\partial F^{\lambda\mu}}{\partial x^\nu}&+&\frac{\partial F^{\nu\lambda}}{\partial x^\mu} =0.
	\end{eqnarray}
\end{subequations}
where $F^{G\left[ \mu\nu\right] }=\frac{1}{2}\left( F^{G\mu\nu}-F^{G\nu\mu}\right)$ is the antisymmetric part of the geometric field. The energy-momentum tensor is generalized to
\begin{equation}\label{EMT2}
	T_{\mu\nu}=\mathcal{F}_{\mu\sigma}\mathcal{F}^{\sigma}\ _\nu-\frac{1}{4}g_{\mu\nu}\mathcal{F}_{\alpha\beta}\mathcal{F}^{\alpha\beta}.
\end{equation}

It should be reminded that the working variables of the Einstein-Maxwell equations are the spacetime metric $\widetilde{g}_{\mu\nu}$ and electromagnetic fields $F_{\mu\nu}$ or $A_\mu$ even the Einstein tensor is expressed in terms of the Finslerian metric $g_{\mu\nu}$.
The spacetime metric contains $10$ degrees of freedom and the electromagnetic field has $6$ degrees of freedom. The total variables are $16$.
The Einstein field equations (\ref{EMeq2}a) contain $10$ independent equations. The generalized Maxwell equations (\ref{EMeq2}b) and (\ref{EMeq2}c) have $8$ independent equations. The coordinate or gauge condition of the metric gives $4$ equations. Thus, there are total $22$ differential equations. On the other hand, the Bianchi identities give constraints. The energy-momentum conservations give $4$ constraints. The current conservation $\partial_\mu J^\mu=0$ implies one constraint and the contraction of the partial derivatives of the source free equation also give one constraint. Consequently, the number of the independent differential equations is $16=22-6$, which is associated with the $16$ working variables $\widetilde{g}_{\mu\nu}$ and $F_{\mu\nu}$.

It should be remarked that the energy-momentum tensor in (\ref{EMT1}) is non symmetric because the geometric field is neither symmetric nor antisymmetric in general. This non symmetric property corresponds to the non symmetric $\Lambda_{\mu\nu}$.\cite{Li2} However, the covariant divergence of the Einstein tensor vanishes implies a novel constraint on the off-diagonal elements of the energy-momentum tensor.

Moreover, the geometric field comes from the geodesic coefficient, which depends only on the Finslerian metric and Finsler function. In other words, the geometric field is coordinate-invariant. Hence the generalized Maxwell equations are also coordinate-invariant.
On the other hand, in the Berwald space, the $hv$ part of the Chern curvature vanishes, which implies that the generalized Maxwell equation is projected on the $hh$ section.

\subsection{Generalized Maxwell equation}\label{subsec4.3}
In order to reveal some basic physics in the Finsler-Randers space, we
give the vector form of the generalized Maxwell equations. The electromagnetic field is expressed in terms of the electric and magnetic fields, $F^{0j}=-E^j$, $F^{j0}=E^j$ and $F^{jk}=-\varepsilon^{jk\ell}B^\ell$. Similarly, we endow the geometric field with the effective electric and magnetic fields, namely effective geometric electric and magnetic fields are defined by,
\begin{equation}\label{FGab1}
	\left[F^{G\mu\nu} \right]\equiv\left(
	\begin{array}{cccc}
		F^{G00} & F^{G01} & F^{G02} & F^{G03} \\
		F^{G10} & F^{G11} & F^{G12} & F^{G13} \\
		F^{G20} & F^{G21} & F^{G22} & F^{G23} \\
		F^{G30} & F^{G31} & F^{G32} & F^{G33} \\
	\end{array}\right):=\left(
	\begin{array}{cccc}
		\mathcal{E}_{00} & \mathcal{E}_{x} & \mathcal{E}_{y} & \mathcal{E}_{z} \\
		\mathcal{E}^T_{x} & \mathcal{B}_{xx} & \mathcal{B}^T_{z} & \mathcal{B}_{y} \\
		\mathcal{E}^T_{y} & \mathcal{B}_{z} & \mathcal{B}_{yy} & \mathcal{B}^T_{x} \\
		\mathcal{E}^T_{z} & \mathcal{B}^T_{y} & \mathcal{B}_{x} & \mathcal{B}_{zz}
	\end{array}\right).
\end{equation}
This definition is equivalent to the vector form of the geometric fields,
\begin{subequations}\label{EEBB1}
	\begin{eqnarray}
		\mathcal{E} &=& \mathcal{E}_x\mathbf{i}+\mathcal{E}_y\mathbf{j}+\mathcal{E}_z\mathbf{k}= F^{G01}\mathbf{i}+F^{G02}\mathbf{j}+F^{G03}\mathbf{k},\\
		\mathcal{E}^T &=& \mathcal{E}^T_x\mathbf{i}+\mathcal{E}^T_y\mathbf{j}+\mathcal{E}^T_z\mathbf{k}= F^{G10}\mathbf{i}+F^{G20}\mathbf{j}+F^{G30}\mathbf{k},\\
		\mathcal{B} &=& \mathcal{B}_x\mathbf{i}+\mathcal{B}_y\mathbf{j}+\mathcal{B}_z\mathbf{k}= F^{G32}\mathbf{i}+F^{G13}\mathbf{j}+F^{G21}\mathbf{k},\\
		\mathcal{B}^T &=& \mathcal{B}^T_x\mathbf{i}+\mathcal{B}^T_y\mathbf{j}+\mathcal{B}^T_z\mathbf{k}= F^{G23}\mathbf{i}+F^{G31}\mathbf{j}+F^{G12}\mathbf{k}.
	\end{eqnarray}
\end{subequations}
Note that the source-free equation in (\ref{EMeq2}c) does not involve the geometric features, which remains the same form as the conventional vector representation.
Consequently, the generalized Maxwell equations (\ref{EMeq2}b) and (\ref{EMeq2}c) can be rewritten as the vector form,
\begin{subequations}\label{EMeq3}
	\begin{eqnarray}
		\nabla\cdot \mathbf{E}
			&-& \frac{1}{2}\mathcal{F}^{\rho\sigma}
		\frac{\partial^2 F^G_{\rho\sigma}}{\partial x^\nu(\partial_\nu A_\mu)} 
		-\frac{1}{4}\left( \frac{\partial \mathcal{F}^{\rho\sigma}}{\partial x^\nu}
		\frac{\partial F^G_{\rho\sigma}}{\partial (\partial_\nu A_\mu)}
		+\frac{\partial \mathcal{F}_{\rho\sigma}}{\partial x^\nu}
		\frac{\partial F^{G\rho\sigma}}{\partial (\partial_\nu A_\mu)}\right)  \\
		&=& \rho+\frac{1}{2}\nabla\cdot \left(\mathcal{E}^T-\mathcal{E}\right),  \\
		\frac{\partial \mathbf{E}}{\partial\tau}-\nabla\times \mathbf{B} 
		&-&\frac{1}{2}\mathcal{F}^{\rho\sigma}
	\frac{\partial^2 F^G_{\rho\sigma}}{\partial x^\nu(\partial_\nu A_\mu)} 
	-\frac{1}{4}\left( \frac{\partial \mathcal{F}^{\rho\sigma}}{\partial x^\nu}
	\frac{\partial F^G_{\rho\sigma}}{\partial (\partial_\nu A_\mu)}
	+\frac{\partial \mathcal{F}_{\rho\sigma}}{\partial x^\nu}
	\frac{\partial F^{G\rho\sigma}}{\partial (\partial_\nu A_\mu)}\right) 	\\	
	&=& \mathbf{J}+\frac{1}{2}\frac{\partial}{\partial\tau}\left(\mathcal{E}-\mathcal{E}^T\right)
		-\frac{1}{2}\nabla\times\left(\mathcal{B}^T-\mathcal{B}\right), \\
		&& \nabla\times\mathbf{E}+\frac{\partial\mathbf{B}}{\partial \tau} = 0,  \\
		&& \nabla\cdot\mathbf{B} = 0,
	\end{eqnarray}
\end{subequations}

It can be seen that the geometric field modifies the generalized Maxwell equations in the Finsler-Randers space. The geometric effects in the left-hand sides of the equations (\ref{EMeq3}) can be regarded as the modification from the geometric field, including the couplings between the electromagnetic field and geometric field as well as their derivatives with respect to the vector potentials and their gradients.
The term $\nabla\cdot \left(\mathcal{E}-\mathcal{E}^T\right)$ in the right-hand sides of the equation (\ref{EMeq3}a) can be interpreted as a geometric source induced by the divergence of the geometric field. The term $\frac{\partial}{\partial\tau}\left(\mathcal{E}-\mathcal{E}^T\right)$
in (\ref{EMeq3}b) can be understood as a geometric current which is an analog with the Dirac electric displacement current.
The term $\nabla\times\left(\mathcal{B}-\mathcal{B}^T\right)$ can be also regarded as a geometric current associated with the vorticity of the geometric field.
When the geometric field vanishes, the generalized Maxwell equations reduce to the standard form.

It should be remarked that the Einstein-Maxwell equations in the Finsler-Randers space involves two geometric structures, $\widetilde{g}_{\mu\nu}$ and $g_{\mu\nu}$. They describe spacetime (or gravity) and dynamics. This formalism provides a unified description of spacetime (or gravity) and dynamics of system.

\subsection{Generalized wave equation}\label{sec4.4}
To explore the wave behavior in the Finsler-Randers space, we consider a simplified situation. In the weak field and low velocity approximation, we ignore the coupling effects of the electromagnetic and geometric field in the generalized Maxwell equations (\ref{EMeq3}). Thus, the generalized Maxwell equations can be rewritten as
\begin{subequations}\label{EMeq4}
	\begin{eqnarray}
		\nabla\cdot \mathbf{E}&=&\rho^G, \\
		-\frac{\partial \mathbf{E}}{\partial\tau}+\nabla\times \mathbf{B}
		&=&\mathbf{J}^G,\\
		\nabla\times\mathbf{E}+\frac{\partial\mathbf{B}}{\partial \tau} &=& 0,  \\
		\nabla\cdot\mathbf{B} &=& 0,
	\end{eqnarray}
\end{subequations}
where
\begin{subequations}\label{RCC2}
	\begin{eqnarray}
		\rho^G & :=& \rho+ \rho_\mathcal{E},\\
		\mathbf{J}^G & :=& \mathbf{J}+\mathbf{J}_\mathcal{E}+\mathbf{J}_\mathcal{B}.
	\end{eqnarray}
\end{subequations}
are the total charges and currents, and
\begin{subequations}\label{RCC0}
	\begin{eqnarray}
		\rho_\mathcal{E} &=& \frac{1}{2}\nabla \cdot (\mathcal{E}^T-\mathcal{E}), \\
		\mathbf{J}_\mathcal{E} &=& \frac{1}{2}\frac{\partial }{\partial\tau}(\mathcal{E}-\mathcal{E}^T), \\
		\mathbf{J}_\mathcal{B} &=& \frac{1}{2}\nabla\times(\mathcal{B}^T-\mathcal{B}) .
	\end{eqnarray}
\end{subequations}

Using the vector calculus techniques, the generalized Maxwell equations (\ref{EMeq4}) can be rewritten as the generalized wave equations,
\begin{subequations}\label{GWqu1}
	\begin{eqnarray}
		\frac{\partial^2\mathbf{E}}{\partial \tau^2}-\nabla^2 \mathbf{E} &=& \frac{\partial \mathbf{J}^G}{\partial \tau}+\nabla \rho^G,\\
		\frac{\partial^2\mathbf{B}}{\partial \tau^2}-\nabla^2 \mathbf{B} &=& -\nabla\times \mathbf{J}^G.
	\end{eqnarray}
\end{subequations}
This generalized wave equations are nonhomogeneous even in the absence of electric charges and currents. The geometric field plays the roles of effective charge and currents to modify the behaviors of electromagnetic wave. Interestingly, the effective charges and currents depend on spacetime geometry or gravity and dynamics. In general, they actually couple with the Einstein field equation (\ref{EST0}a). These general Einstein-Maxwell equations are very difficult to be solved analytically. However, when gravity can be regarded as spacetime background, the nonhomogeneous terms play a source role in the generalized wave equation. These spacetime background sources could provide a physical scenario of cosmological microwave radiation.

On the other hand, we may suppose the nonhomogeneous terms vanishing as a particular gauge conditions of the electromagnetic fields, namely
\begin{subequations}\label{GWqu2}
	\begin{eqnarray}
		\frac{\partial \mathbf{J}^G}{\partial \tau}+\nabla \rho^G &=& 0,\\
		\nabla\times \mathbf{J}^G &=& 0.
	\end{eqnarray}
\end{subequations}
These gauge conditions (\ref{GWqu2}) can be also understood as the effective current conservations in the Finsler-Randers space. Consequently,
the generalized wave equations reduce to those in vacuum,
\begin{subequations}
	\begin{eqnarray}
		\frac{\partial^2\mathbf{E}}{\partial \tau^2}-\nabla^2 \mathbf{E} &=& 0, \\
		\frac{\partial^2\mathbf{B}}{\partial \tau^2}-\nabla^2 \mathbf{B} &=& 0.
	\end{eqnarray}
\end{subequations}
In this gauge condition, we obtain a generalized current conservation law and the plane wave solution of the generalized Maxwell equation in the Finsler-Randers space.
How to experimentally test this result is still a challenging issue. It will be expected to be studied further.

\section{Typical cases}\label{sec5}
Let us investigate a few typical cases to reveal some basic physical behaviors of the system in the Finsler-Randers space.

\subsection{Flat spacetime $\widetilde{g}_{\mu\nu}=\eta_{\mu\nu}$}\label{subsec5.1}
Firstly, we consider the flat spacetime $\widetilde{g}_{\mu\nu}=\eta_{\mu\nu}$, where $\eta_{\mu\nu}$ is the Minkowski metric. %Note that the effective mass, $m^E_{\mu\nu}=2g_{\mu\nu}/\mathcal{L}$,
the effective mass is given by
\begin{equation}\label{FMT2}
	m^E_{\mu\nu}=\frac{1}{\alpha}\left(m^2\eta_{\mu\nu}-\ell_\mu \ell_\nu\right)+(\ell_\mu+b_\mu)(\ell_\nu+b_\nu).
\end{equation}
Interestingly, the effective mass depends on the velocity and electromagnetic fields. This behavior is similar to electrons moving in semiconductor. \cite{Mermin}
Since in the flat spacetime, $Q_{\nu\lambda\sigma}=M_{\nu\lambda\sigma}=0$, the geodesic coefficient is simplified to
\begin{equation}\label{EMTa}
	G^\mu=\frac{1}{4}g^{\mu\nu}\left(-2\alpha F_{\nu\sigma}+S_{\nu\lambda\sigma}v^\lambda\right)y^\sigma.
\end{equation}
Consequently, the total effective force reduces to
\begin{equation}\label{fff1}
	f_{\mu}=\left(1-\frac{\beta}{\mathcal{L}}\right)F_{\mu\nu}y^\nu-\frac{1}{2\mathcal{L}}
	S_{\mu\nu\sigma}y^\sigma y^\nu,
\end{equation}
and its corresponding field can be given by
\begin{equation}\label{fff2}
	\mathcal{F}_{\mu\nu}=\left(1-\frac{\beta}{\mathcal{L}}\right)F_{\mu\nu}-\frac{1}{2\mathcal{L}}S_{\mu\nu\sigma}y^\sigma.
\end{equation}
Thus, the geodesic equation can be expressed as
\begin{equation}\label{GEq2}
	\dot{y}^\mu+\frac{1}{2}g^{\mu\nu}\left(-2e\alpha F_{\nu\sigma}+S_{\nu\lambda\sigma}y^\lambda\right)y^\sigma=0.
\end{equation}
It can be seen that the Finsler geometric feature modifies the Lorentz force. The component coupling of the electromagnetic field induces an additional effective force and field, which drives the geodesics departure from those in the Euclidian and Riemannian geometries even in the flat spacetime.

\subsection{Vacuum $A_\mu=0$}\label{subsec5.2}
The second simplified case is that the electromagnetic field vanishes, $A_\mu=0$, namely, the Finsler-Randers space reduces to the Finsler-Riemann space $\beta=0$. The effective mass reduces to
\begin{equation}\label{FMT3}
	m^E_{\mu\nu}=\frac{1}{\alpha}\left(m^2\widetilde{g}_{\mu\nu}+\ell_\mu \ell_\nu\right)+\ell_\mu\ell_\nu.
\end{equation}
The effective mass still depends on the velocity field. Since $F_{\nu\sigma}=M_{\nu\lambda\sigma}=S_{\nu\lambda\sigma}=0$ when $A_\mu$ vanishes,
the geodesic coefficient is simplified to
\begin{equation}\label{EMT3}
	G^\mu=\frac{1}{4}g^{\mu\nu}Q_{\nu\lambda\sigma}y^\lambda y^\sigma.
\end{equation}
The effective force becomes
\begin{equation}\label{fff3}
	f_{\mu}=-\frac{1}{2\mathcal{L}}Q_{\mu\nu\sigma}y^\sigma y^\nu,
\end{equation}
and its corresponding effective field reduces to
\begin{equation}\label{fff4}
	F^G_{\mu\nu}=-\frac{1}{2\mathcal{L}}Q_{\mu\nu\sigma}y^\sigma.
\end{equation}
Consequently, the geodesic equation is obtained
\begin{equation}\label{GEq3}
	\dot{y}^\mu+\frac{1}{2}g^{\mu\nu}Q_{\nu\lambda\sigma}y^\lambda y^\sigma=0.
\end{equation}
Interestingly, the curved spacetime generates an effective force and field even in vacuum. In other words, particles move with acceleration driven by the geometric force or curved spacetime.

When we ignore the coupling terms in the weak field and low velocity approximations,
the generalized Maxwell equations (\ref{EMeq4}b) can be also expressed as
\begin{subequations}
	\begin{eqnarray}
		\nabla \cdot (\mathcal{E}^T-\mathcal{E}) &=& 0, \\
		\frac{\partial}{\partial \tau} (\mathcal{E}^T-\mathcal{E})+\nabla\times (\mathcal{B}-\mathcal{B}^T) &=& 0.
	\end{eqnarray}
\end{subequations}
On the other hand, the generalized Maxwell equations (\ref{EMeq2}b) can be reduced to $\frac{\partial F^{G[\nu\mu]}}{\partial x^\nu}=0$.
Using the geometric field in (\ref{fff4}), the Finslerian function is constrained on the affine sphere $\mathcal{L}(x,y)=1$ for convenience,
the generalized Maxwell equation can be expressed as
\begin{equation}\label{ggg1}
	y^\kappa g^{\nu\rho}g^{\mu\sigma}\left(\frac{\partial^2 \widetilde{g}_{\rho\kappa}}{\partial x^\nu\partial x^\sigma}
	-\frac{\partial^2 \widetilde{g}_{\sigma\kappa}}{\partial x^\nu\partial x^\rho}\right)
	+y^\kappa \left(\frac{\partial \widetilde{g}_{\rho\kappa}}{\partial x^\sigma}
	-\frac{\partial \widetilde{g}_{\sigma\kappa}}{\partial x^\rho}\right)\frac{\partial g^{\nu\rho}g^{\mu\sigma}}{\partial x^\nu}=0.
\end{equation}
Interestingly, the generalized Maxwell equation in the Randers space is a nonlinear second-order differential equation with the coupling between the Finslerian and spacetime metrics even in vacuum. In other words, the interplay of gravity and dynamics emerges in the Randers space. This result uncovers an intrinsic geometrodynamics in Finsler geometry.

\subsection{Berwald space}\label{subsec5.3}
The third case is that we consider the system in the subspace of the Finsler space where the $hv$ part of the Chern curvature vanishes. In this case, the geodesic coefficient is given by,
\begin{equation}\label{GGb}
	G^\lambda=\frac{1}{2}\gamma^\lambda\ _{\mu\nu} y^\mu y^\nu,
\end{equation}
where
\begin{equation}\label{Gama0}
	\gamma^\lambda\ _{\mu\nu}=\frac{1}{2}g^{\lambda \sigma}\left(\frac{\partial g_{\sigma\nu}}{\partial x^\mu }
	+\frac{\partial g_{\mu\sigma}}{\partial x^\nu}-\frac{\partial g_{\mu\nu}}{\partial x^\sigma}\right).
\end{equation}
is the Christoffel symbol of the Finsler geometry. The Berwald condition (\ref{BbB0}) $B_{j|k}=0$ leads to the constraint on electromagnetic field,
\begin{equation}\label{Bb1}
	\frac{\partial A_\mu}{\partial x^\nu}-A_\sigma\widetilde{\gamma}^\sigma\ _{\mu\nu}=0.
\end{equation}
The geodesic coefficient can be expressed as two equivalent forms,\cite{Bao} (See Appendix B.3)
\begin{equation}\label{GG3}
	2G^\lambda=\gamma^\lambda\ _{\mu\nu} y^\mu y^\nu=\widetilde{\gamma}^\lambda\ _{\mu\nu} y^\mu y^\nu,
\end{equation}
where
\begin{equation}\label{Gama1}
	\widetilde{\gamma}^i\ _{jk} = \frac{1}{2}\widetilde{g}^{i \ell}\left(\frac{\partial \widetilde{g}_{\ell k}}{\partial x^j}
	+\frac{\partial \widetilde{g}_{j\ell}}{\partial x^k}-\frac{\partial \widetilde{g}_{jk}}{\partial x^\ell}\right)
\end{equation}
is the Christoffel symbol of the spacetime geometry, namely the spacetime connection coefficient.
Consequently, the effective force can be expressed as
\begin{equation}\label{fff5}
	f_{\mu} =-\frac{1}{\mathcal{L}}g_{\mu\nu}\gamma^\nu\ _{\lambda\sigma} y^\lambda y^\sigma
	=-\frac{1}{\mathcal{L}}g_{\mu\nu}\widetilde{\gamma}^\nu\ _{\lambda\sigma} y^\lambda y^\sigma
\end{equation}
and their corresponding fields can be given by
\begin{equation}\label{ffd6}
	\mathcal{F}_{\mu\nu} =-\frac{1}{\mathcal{L}}g_{\mu\sigma}\gamma^\sigma\ _{\lambda\nu} y^\lambda
	=-\frac{1}{\mathcal{L}}g_{\mu\sigma}\widetilde{\gamma}^\sigma\ _{\lambda\nu} y^\lambda.
\end{equation}
The effective force and its corresponding field show two equivalent forms, which correspond to two geometric structures associated with spacetime and dynamics.

In the physical viewpoint, the condition in (\ref{Bb1})
ensures the Chern connection coefficient to be independent of $y$. 
Interestingly, note that $B_{j|k}-B_{k|j}$ gives the electromagnetic field, which can be expressed as 
\begin{equation}\label{FTS1}
	F_{\mu\nu}=A_\sigma \left( \widetilde{\gamma}^\sigma\ _{\nu\mu}-\widetilde{\gamma}^\sigma\ _{\mu\nu}\right) =0,
\end{equation}
because the spacetime is assumed torsionless, namely
Christoffel symbol $\widetilde{\gamma}^\lambda_{\nu\mu}$ is symmetric.
In the Mathematical viewpoint, the Berwald condition is referred to as the Berwald structure,\cite{Bao} which deviates from the Riemannian structure, in whcih
the $hv$-part of the Chern-Riemann curvature vanishes, namely $\frac{\delta}{\delta x^\mu}\rightarrow \frac{\partial}{\partial x^\mu}$ in the formulation of the
the Chern-Riemann curvature.

It should be remarked that torsion free means electromagnetic field vanishes, but it does not imply magnetic potential disappearing. The source Maxwell equation (\ref{EMeq2}b) can be expressed as

\begin{equation}\label{MeqB}
		\frac{\partial F^{G[\mu\nu]}}{\partial x^\nu}-\frac{1}{2} F^{G\rho\sigma}
		\frac{\partial^2 F^G_{\rho\sigma}}{\partial x^\nu(\partial_\nu A_\mu)} \\
		-\frac{1}{4}\left( \frac{\partial F^{G\rho\sigma}}{\partial x^\nu}
		\frac{\partial F^G_{\rho\sigma}}{\partial (\partial_\nu A_\mu)}
		+\frac{\partial F^G_{\rho\sigma}}{\partial x^\nu}
		\frac{\partial F^{G\rho\sigma}}{\partial (\partial_\nu A_\mu)}\right) 
		=J^\mu.
\end{equation}

Interestingly, it can be seen that charge and current can drive the effective geometric field in the Berwald space. Moreover, $A_\mu \neq 0$ with $F_{\mu\nu}=0$ could lead to some topological phenomena.

	% =========================
	\section{What physics are these geometric formulations?}
	% =========================

The geometric representation of physical systems provides a novel insight to the mathematical structures behind physical phenomena. Different geometric structures as different measurement rules or references give different angles to project physical observations.

What physics do we find from these geometric formulations? Firstly,
the generalized Einstein-Maxwell equations provide a unified description of gravity and dynamics. The Finslerian metric modifies spacetime geometry, which gives interplay between dynamics and gravity. The geodesic equation in the Finsler-Randers space gives effective (or geometric) fields as an analog with a generalized electromagnetic field. We find that there are three-type geometric fields. One depends on the curved spacetime, $Q_{\mu\nu\sigma}$. The second depends on the couplings of the electromagnetic fields with velocity fields, $S_{\mu\nu\sigma}$. The third depends on the coupling between the curved spacetime and electromagnetic field, $M_{\mu\nu\sigma}$. These geometric fields provide some hints to understand the physical mechanisms of some unsolved puzzles, such as axion, dark matter, dark energy and gravitational wave. \cite{Hooft,Sikivie,Wilczek,Luca1,Luca2,David}

Axion as a pseudo-particle was proposed to solve the strong (Charge-Parity)CP problem in QCD.\cite{Peccei,Hooft}
The axion is an interaction between gravity and electromagnetic field with zero electric charge and zero spin. One possible candidate of axion is a novel particle appearing as vacuum fluctuation with CP violation in particle physics.\cite{Peccei,Hooft} The second possibility is a cold dark matter in cosmology.\cite{Sikivie,Wilczek}
The internal symmetry (duality) of electromagnetic field could be spontaneously broken by a pseudoscalar axion-link field, \cite{Wilczek,Luca1}
\begin{equation}\label{AxF1}
	\mathcal{L}_\theta=-\frac{\kappa}{\mu_0c}\theta(x) \mathbf{E}\cdot \mathbf{B},
\end{equation}
where $\theta(x)$ is a pseudoscalar field and $\kappa$ is the coupling constant. This axion-electromagnetic coupling could be related to the effective or geometric fields $S_{\mu\nu\sigma}$ and $M_{\mu\nu\sigma}$ in the Finsler-Randers space. In particular, the couplings between electric, magnetic and geometric fields emerge in the generalized Maxwell equations. In curved spacetime, the axion-electromagnetic field coupled with gravity provides a physical scenario to detect the gravitational wave. The geometric field $M_{\mu\nu\sigma}$ shows the coupling of the electromagnetic field and gravity. This give some hints to understand axion-gravitational wave. \cite{Luca1,Luca2,David}
Recently, a few experimental schemes were proposed to detect the axion signature from the high frequency gravitational wave.\cite{Luca1,Luca2}
These formulations can be applied to explore possibilities of the physics mechanisms of axion on the QCD vacuum, dark matter and axion gravitational wave.

More interestingly, the geometric part $\Lambda_{\mu\nu}$ in the Einstein tensor can be interpreted as a geometric feature of the cosmological constant or dark energy in the Finsler-Randers space. In other words, These results give us a novel view to dark energy and inspire some hints to detect experimentally dark energy based on Finsler geometry. 

\section{Conclusions and outlooks}\label{sec6}
In summary, we construct a unified framework of geometrodynamics based on the Finsler geometry.
The Lagrangian of electron in electromagnetic fields plays the role of the Finsler function with the Finsler-Randers structure.
In the weak field and low velocity approximations, the Finslerian metric modifies spacetime geometry, which gives geodesic equation from the Euler-Lagrange equation.
The Newton-type equation gives the effective mass, force and geometric fields. These geometric fields allow us to construct the generalized Maxwell equations in the Finsler-Randers space. Using the Chern connection, Ricci and scalar tensors, we construct the Einstein tensor and field equations. Consequently, we obtain the generalized Einstein-Maxwell equation in the Finsler-Randers space.

What do we find beyond the Riemannian gravity from this geometrodynamics? The Einstein tensor contains an additional geometric term. This geometric feature provides some hints to understand dark energy as a geometric view of cosmological constant. We find that there are couplings between the electric, magnetic and geometric fields in the generalized Maxwell and wave equations. The divergence and vorticity of the geometric field play effective charges and currents, which
provide a geometric view to understand the Dirac (or magnetic) monopole, axion in CP problem and dark matter as well as the cosmic background radiation.
A few simplified cases, such as flat spacetime, vacuum and Berwald space, reveal some basic physical properties of the Finsler-Randers space. In particular, we find that electromagnetic field vanishes in the Berwald space, but there still exists the magnetic potential. The interplay of two geometries, spacetime (or gravity) and dynamics emerges in the Randers space.

The emergence of the additional geometric term $\Lambda_{\mu\nu}$ in the Einstein tensor could reveal two physical scenarios for understanding some puzzles. This geometric term can be interpreted as a geometric description of cosmological constant or dark energy. On the other hand, it implies that the energy-momentum contains some novel structures associated with the mathematical structure of this geometric term. \cite{Li2}  The non symmetric energy-momentum tensor implies a novel constraint of the off-diagonal elements in the energy-momentum tensor or the violation of Lorentz invariance.\cite{Neil,Schreck,Li2}

In fact, this framework of geometrodynamics is far from completeness. However, more importantly, it simulates novel thoughts on deeper geometric structures behind physics. Is geometric structure reality or measurement rules? How do we implement different geometric measurement? Can we observe the Finslerian features in laboratory? Actually, many issues are worth studying further. For example, what are the fundamental physical properties of the geometric fields? Are they observable? In general, observations depend on connections and local frames in curved spacetimes. These formulations can be also applied to more practical models, such as axion, Kerr black hole, cosmological model, and Hall effect in cosmology.\cite{Pierre,Jacek}

Mathematically, we give a real physical application in Finsler geometry, especially in the Randers space with or without Berwald structure. These results tell us what physical observables or phenomena emergence from geometric structures even though our model is associated with approximations and assumptions. These physical predictions could inspire some clues to unlock some novel mathematical structures.

\section*{Declarations}

%Some journals require declarations to be submitted in a standardised format. Please check the Instructions for Authors of the journal to which you are submitting to see if you need to complete this section. If yes, your manuscript must contain the following sections under the heading `Declarations':

\begin{itemize}
	\item Funding: Not applicable
	\item Conflict of interest/Competing interests: %(check journal-specific guidelines for which heading to use)
	The authors declare that they have no conflict of interest and competing interests.
	\item Ethics approval: The authors declare that they have upheld the integrity of the scientific record.
	\item Consent to participate: The authors give their consent for publication of this article.
	\item Consent for publication: The authors give their consent for publication of this article.
	\item Availability of data and materials: Not applicable
	\item Code availability: Not applicable
	\item Authors' contributions: The authors contributed equally to this work, and approved the final manuscript.
\end{itemize}

\section{Appendices}

\subsection{Preliminaries on Finsler geometry}\label{secA1}
\subsubsection{Chern connections}\label{subsecA1}
Let $(M,F)$ be a Finsler manifold. The pulled-back bundle $\pi^*TM$ admits a unique linear connection called Chern connection.
The connection forms are characterized by two structure equations,\cite{Bao}
\begin{subequations}\label{CC1}
	\begin{eqnarray}
		d\omega^\mu-\omega^\nu\wedge\omega_\nu\ ^\mu &=& 0, \quad \textrm{Torsion freeness}\\
		dg_{\mu \nu}-g_{\rho\nu}\omega_\nu\ ^\rho-g_{\mu\rho}\omega_\nu\ ^\rho &=&2C_{\mu\nu\rho}\delta y^{\rho} \quad \textrm{Almost g-compatibility},
	\end{eqnarray}
\end{subequations}
where $\omega^\mu :=dx^\mu$ and
$\delta y^\mu:=dy^\mu+N^\mu\ _{\nu}dx^\nu$ with $N^\mu\ _\nu :=\frac{\partial G^\mu}{\partial y^\nu}$ being the Cartan nonlinear connection.
\begin{equation}\label{CTS1}
	C_{\mu\nu\rho}=\frac{1}{4}\frac{\partial^3F^2}{\partial y^\mu\partial y^\nu\partial y^\rho}
\end{equation}
is the Cartan tensor, which describes the deviation from the Riemanian structure.

The torsion freeness (\ref{CC1}a) means the absence of $dy^\mu$ terms in $\omega_\nu\ ^\mu$, namely
\begin{equation}\label{CC2}
	\omega_\nu\ ^\mu=\Gamma^\mu\ _{\nu\rho}dx^\rho,
\end{equation}
where $\Gamma^\mu\ _{\nu\rho}=\Gamma^\mu\ _{\rho\nu}$ is called the Chern connection coefficient.
The almost g-compatibility (\ref{CC1}b) implies
\begin{equation}\label{CC3}
	\frac{\partial g_{\mu\nu}}{\partial x^\sigma}-2C_{\mu\nu\rho}N^\rho\ _\sigma=g_{\rho\nu}\Gamma^\rho\ _{\mu\sigma}+g_{\rho\mu}\Gamma^\rho\ _{\nu\sigma}.
\end{equation}
By using the so-called Christoffel's trick, the Chern connection coefficients can be expressed as,\cite{Bao}
\begin{equation}\label{CC4}
	\Gamma^\mu\ _{\nu\rho}=\frac{1}{2}g^{\mu\sigma}\left(\frac{\delta g_{\sigma\rho}}{\delta x^{\nu}}
	+\frac{\delta g_{\sigma\nu}}{\delta x^{\rho}}-\frac{\delta g_{\nu\rho}}{\delta x^{\sigma}}\right),
\end{equation}
where $\frac{\delta}{\delta x^\mu}:=\frac{\partial}{\partial x^\mu}-N^\sigma\ _\mu\frac{\partial}{\partial y^\sigma}$.

%It should be reminded that the Chern connection is defined on $\pi^*TM$ that is a vector bundle over the slit tangent bundle $TM_0$,
%but the Levi-Civita connection is usually defined on the manifold $M$.

\subsubsection{Covariant derivative and differential}\label{subsecA2}
Let $T:=T^\nu\ ^\mu \frac{\partial}{\partial x^\nu}\otimes dx^\mu$ be an arbitrary smooth local section of $\pi^*TM\otimes  \pi^*T^*M$. It is a tensor field with
rank $(1,1)$ on $TM_0$. Its covariant differential is defined by
\begin{equation}\label{TT0}
	\nabla T:=(\nabla T)^\nu\ _\mu \frac{\partial}{\partial x^\nu}\otimes dx^\mu,
\end{equation}
where
\begin{equation}\label{TT1}
	(\nabla T)^\nu\ _\mu:=dT^\nu\ _\mu+T^\kappa\ _\mu\omega_\kappa\ ^\nu-T^\nu\ _\kappa\omega_\mu\ ^\kappa.
\end{equation}
Note that $(\nabla T)^\nu\ _\mu$ is 1-form on $TM_0$. They can be expanded in terms of the natural basis $\left\{dx^\lambda, \delta y^\lambda\right\}$,
\begin{equation}\label{TT2}
	(\nabla T)^\nu\ _\mu=T^\nu\ _{\mu|\lambda}dx^\lambda+T^\kappa\ _{\mu;\lambda}\delta y^\lambda.
\end{equation}
In order to obtain formulas for the coefficients, we evaluate equation (\ref{TT2}) on each individual member of the dual basis
$\left\{\frac{\delta}{\delta x^\lambda},\frac{\partial}{\partial y^\lambda}\right\}$. Using $\omega_\nu\ ^\mu=\Gamma^\mu\ _{\nu\lambda}dx^\lambda$, and
\begin{subequations}\label{LCD1}
	\begin{eqnarray}
		\nabla _{\frac{\partial}{\partial x^\rho}}\frac{\partial}{\partial x^\mu} &=&\Gamma^\nu\ _{\mu\rho}\frac{\partial}{\partial x^\nu},\\
		\nabla _{\frac{\partial}{\partial x^\rho}}dx^{\nu}&=&-\Gamma^\nu\ _{\mu\rho}dx^{\mu},
	\end{eqnarray}
\end{subequations}
we obtain
\begin{subequations}\label{TT3}
	\begin{eqnarray}
		T^\nu\ _{\mu|\lambda}&=&\left(\nabla_{\frac{\delta}{\delta x^\lambda}}T\right)^\nu\ _\mu=\frac{\delta T^\nu\ _\mu}{\delta x^\lambda}+T^\kappa\ _\mu\Gamma^\nu\ _{\kappa\lambda}-T^\nu\ _\kappa\Gamma^\kappa\ _{\mu\lambda}, \\
		T^\nu\ _{\mu;\lambda}&=&\left(\nabla_{\frac{\partial}{\partial y^\lambda}}T\right)^\nu\ _\mu=\frac{\partial T^\nu\ _\mu}{\partial y^\lambda}.
	\end{eqnarray}
\end{subequations}
It should be remarked that the covariant derivative in the natural basis,
\begin{itemize}
	\item the horizontal covariant derivative $T^\nu\ _{\mu|\lambda}$ is comprised of a horizontal directional derivative $\frac{\delta T^\nu\ _\mu}{\delta x^\lambda}$ and
	correction terms.
	\item the vertical covariant derivative $T^\nu\ _{\mu;\lambda}$ consists merely of a homogenized partial derivative. There are no correction terms.
\end{itemize}
Here we list some typical examples of covariant derivative to some typical objects.
For the Chern connection,
\begin{equation}\label{CGCT1}
	(\nabla g)_{\mu\nu}=dg_{\mu\nu}-g_{\kappa\nu} \omega_\mu\ ^\kappa-g_{\mu\kappa}\omega_\nu\ ^\kappa = 2C_{\mu\nu\lambda}\delta y^\lambda,
\end{equation}
which turns out
\begin{subequations}\label{LCD6}
	\begin{eqnarray}
		g_{\mu\nu|\rho}&=&0, \quad g^{\mu\nu}\ _{|\rho}=0,\\
		g_{\mu\nu;\rho}&=&2C_{\mu\nu\rho}, \quad g^{\mu\nu}\ _{;\rho}=-2C^{\mu\nu}\ _\rho,
	\end{eqnarray}
\end{subequations}
where we use $\left(g^{\mu\sigma}g_{\sigma\nu}\right)_{|\rho}=0$ and $\left(g^{\mu\sigma}g_{\sigma\nu}\right)_{;\rho}=0$.

It implies that the fundamental tensor is covariantly constant along horizontal direction, but its vertical derivatives are proportional to the Cartan tensor.
The horizontal covariant derivative $T^\mu\ _{\nu|\rho}$ depends on the choice of the connection on $\pi^*TM$,
while the vertical covariant derivative $T^\mu\ _{\nu;\rho}$
consists of partial derivative for $y$. The covariant derivatives for higher rank tensor fields are similar.
Recalling the second structural equation (\ref{CC1}b) for the Chern connection, we have\cite{Bao}

\subsubsection{Ricci scalar and Ricci curvature}\label{subsecA3}
Using (\ref{CGCT1}), the torsion-free Chern-Riemann curvature can be expressed as\cite{Bao}
\begin{equation}\label{RCCT1}
	R_\mu\ ^\lambda\ _{\sigma\nu}=\frac{\delta \Gamma^\lambda\ _{\mu\nu}}{\delta x^\sigma}- \frac{\delta \Gamma^\lambda\ _{\mu\sigma}}{\delta x^\nu}
	+\Gamma^\lambda\ _{\mu \rho}\Gamma^\rho\ _{\mu\nu}-\Gamma^\lambda\ _{\nu \rho}\Gamma^\rho\ _{\mu\sigma}.
\end{equation}
Actually, the Chern-Riemanian curvature can be expressed equivalently in terms of the geodesic coefficient.
By doing exterior differential of (\ref{CC1}b), we have
\begin{equation}\label{BBIT0}
	g_{\mu\kappa}\Omega^\kappa_{\nu}+g_{\kappa\nu}\Omega^\kappa_{\mu}
	=-2\left(C_{\mu\nu\kappa|\lambda}\omega^\lambda+C_{\mu\nu\kappa|\lambda}\delta y^\lambda\right)\wedge \delta y^\lambda-2C_{\mu\nu\kappa}\Omega^\kappa_\lambda y^\lambda.
\end{equation}
and note that $R_{\mu\nu\kappa\lambda}=R_\mu\ ^\sigma_{\kappa\lambda}g_{\sigma\nu}$, we have
\begin{subequations}
	\begin{eqnarray}
		R_{\mu\nu\kappa\lambda}+R_{\nu\mu\kappa\lambda} &=& -2C_{\mu\nu\sigma}R^\sigma_{\kappa\lambda}, \\
		R^\mu\ _{\kappa\lambda} &=& y^\sigma R_\sigma\ ^\mu\ _{\kappa\lambda}.
	\end{eqnarray}
\end{subequations}
Using (\ref{RCCT1}) and $\Gamma^\mu_{\sigma\nu}y^\sigma=N^\mu_\nu$,\cite{Shen} we have
\begin{equation}\label{RNT1}
	R^\mu_{\nu\kappa}=\frac{\partial N^\mu_\kappa}{\partial x^\nu}- \frac{\partial N^\mu_\nu}{\partial x^\kappa}
	+N^\sigma_\kappa\frac{\partial N^\mu_\nu}{\partial y^\sigma}-N^\sigma_\nu\frac{\partial N^\mu_\kappa}{\partial y^\sigma}.
\end{equation}
Further using the contraction $y^\kappa R^\mu_{\nu\kappa}$ and $N^\mu_\nu=\frac{\partial G^\mu}{\partial y^\nu}$, the Ricci curvature can be expressed as
\begin{equation}\label{Ricc1}
	R^\mu\ _{\nu}=2\frac{\partial G^\mu}{\partial x^\nu}- y^\kappa\frac{\partial^2 G^\mu}{\partial x^\kappa\partial y^\nu}
	+2G^\kappa\frac{\partial^2 G^\mu}{\partial y^\kappa\partial y^\nu}
	-\frac{\partial G^\mu}{\partial y^\kappa}\frac{\partial G^\kappa}{\partial y^\nu}.
\end{equation}
Thus, the Ricci scalar can be given by the contraction of the index, $Ric=R^\mu\ _\mu$, which is equivalent to
\begin{equation}\label{Ricc2}
	Ric := y^\rho R_\rho\ ^\mu\ _{\mu\sigma} y^\sigma.
\end{equation}
The Ricci curvature tensor is defined by \cite{Bao,Shen}
\begin{equation}\label{RiccT}
	Ric_{\mu\nu}:=\frac{1}{2}\frac{\partial^2 Ric}{\partial y^\mu \partial y^\nu}.
\end{equation}
The Ricci scalar and curvature tensor obey $Ric = y^\mu Ric_{\mu\nu}y^\nu$ and they are compatible with the general Finsler structure.\cite{Bao}

\subsubsection{Chern connection coefficient}
The Chern connection coefficient is defined by
\begin{equation}\label{GGama}
	\Gamma^\lambda\ _{\mu\nu}:=\frac{1}{2}g^{\lambda \sigma}\left(\frac{\delta g_{\sigma\nu}}{\delta x^\mu }
	+\frac{\delta g_{\mu\sigma}}{\delta x^\nu}-\frac{\delta g_{\mu\nu}}{\delta x^\sigma}\right),
\end{equation}
where
\begin{equation}\label{delta}
	\frac{\delta}{\delta x^\mu}:=\frac{\partial}{\partial x^\mu}-N^\sigma\ _\mu\frac{\partial}{\partial y^\sigma}.
\end{equation}
Using the nonlinear connection $N^\mu_\nu$, the Chern connection coefficient can be expressed as
\begin{equation}\label{GGama2}
	\Gamma^\lambda\ _{\mu\nu}=\gamma^\lambda\ _{\mu\nu} -\frac{1}{2}g^{\lambda \sigma}\left(N^\alpha_\mu\frac{\partial g_{\sigma\nu}}{\partial y^\alpha }
	+N^\alpha_\nu\frac{\partial g_{\mu\sigma}}{\partial y^\nu}-N^\alpha_\sigma\frac{\partial g_{\mu\nu}}{\partial y^\alpha}\right).
\end{equation}

\subsection{Calculation notes}

\subsubsection{Finsler metric}
The Lagrangian of an electron in the electromagnetic field in (1) can be rewritten as the Finslerian function in the Randers spaces,\cite{Bao,Miron,Sabau,Jackson,Cheng}
\begin{equation}\label{FS1}
	\mathcal{L}\left(x,y \right)=\alpha(x,y)+\beta(x,y),
\end{equation}
where
\begin{subequations}\label{FS1a}
	\begin{eqnarray}
		\alpha(x,y) &=&  m\widetilde{\alpha}, \quad \widetilde{\alpha}=\sqrt{\widetilde{g}_{\mu\nu} y^\mu y^\nu},\\
		\beta (x,y) &=& b_\mu y^\mu, \quad b_\mu\equiv eA_\mu.
	\end{eqnarray}
\end{subequations}
The definition of the fundamental metric $g_{\mu\nu}$ in (\ref{FMT1}) can be rewritten as,
\begin{equation}\label{FS2}
	g_{\mu\nu}=\mathcal{L}\frac{\partial^2 \mathcal{L}}{\partial y^\mu\partial y^{\nu}}+\frac{\partial \mathcal{L}}{\partial y^\mu}\frac{\partial \mathcal{L}}{\partial y^\nu}.
\end{equation}
Using the Lagrangian in (\ref{FS1}) and introducing $\ell_\mu:=\frac{\partial\alpha}{\partial y^\mu}$, we have
\begin{subequations}\label{FS3}
	\begin{eqnarray}
		\frac{\partial \mathcal{L}}{\partial y^\mu} &=& \ell_\mu+b_\mu, \\
		\frac{\partial^2 \mathcal{L}}{\partial y^\mu\partial y^{\nu}}&=& \frac{\partial \ell_\mu}{\partial y^\nu}.
	\end{eqnarray}
\end{subequations}
Thus, the fundamental metric can be obtained as
\begin{equation}\label{FS4}
	g_{\mu\nu}=\frac{\mathcal{L}}{2\alpha}\left(m^2\widetilde{g}_{\mu \nu}-\ell_{\mu}\ell_{\nu} \right)+\left(\ell_\mu+b_\mu\right)\left(\ell_\nu+b_\nu\right).
\end{equation}
It can be verified that $g_{\mu\nu}$ is non degenerate in the timelike region $v<c$.
Consequently, there exists the inverse of the Finslerian metric. Using the following mathematical theorem,\cite{Cheng} we can calculate the inverse of the Finslerian metric.

\textbf{Theorem}: Let $[g_{ij}]$ and $[A_{ij}]$ be two $n\times n$ symmetric matrices and $B=[B_i]$ be an $n$-dimensional vector, which satisfy
\begin{equation}\label{gg1}
	g_{ij}=A_{ij}+\lambda B_iB_j,
\end{equation}
where $\lambda$ is a constant. Then, the determinant of $g_{ij}$ is given by
\begin{equation}\label{gg2}
	\det(g_{ij})=(1+\lambda B^2)\det(A_{ij}).
\end{equation}
For the positive definite $[A_{ij}]$ there exists $[A_{ij}]^{-1}=[A^{ij}]$ and $1+\lambda B^2\neq 0$. Then $[g_{ij}]$ is invertible, which is given by
\begin{equation}\label{gg3}
	g^{ij}=A^{ij}-\frac{\lambda}{1+\lambda B^2} B^iB^j,
\end{equation}
where $B^i=A^{ij}B_j$ and $B=\sqrt{A^{ij}B_iB_j}$. The proof of this theorem is in the book. \cite{Cheng}

Using this theorem, we can obtain the inverse of the fundamental metric, \cite{Bao,Cheng,Zhou}
\begin{equation}\label{FM1}
	g^{\mu\nu}=\frac{m^2\alpha}{\mathcal{L}}\widetilde{g}^{\mu\nu}-\frac{\alpha}{\mathcal{L}^2}(b^\mu y^\nu+b^\nu y^\mu)
	+\frac{b^2\alpha+\beta}{\mathcal{L}^3}y^\mu y^\nu.
\end{equation}

\subsubsection{Geodesic coefficient}
Based on the fundamental function (\ref{FS1}) we have
\begin{subequations}\label{DDF1}
	\begin{eqnarray}
		\frac{\partial \alpha}{\partial x^\mu} &=& \frac{m}{2\widetilde{\alpha}}\frac{\partial \widetilde{g}_{ab}}{\partial x^\mu}y^a y^b ,\\
		\frac{\partial \beta}{\partial x^\mu} &=& \frac{\partial b_a}{\partial x^\mu}y^a,\\
		\frac{\partial \alpha}{\partial y^\mu} &=& \ell_\mu,  \quad \frac{\partial \beta}{\partial y^\mu} = b_\mu,
	\end{eqnarray}
\end{subequations}
where $\ell_\mu=m\frac{g_{\mu c}v^c}{\widetilde{\alpha}}$. Similarly, we have
\begin{subequations}\label{DDF2}
	\begin{eqnarray}
		\frac{\partial \alpha^2}{\partial x^\mu} &=& m^2\frac{\partial \widetilde{g}_{ab}}{\partial x^\mu}y^a y^b ,\\
		\frac{\partial \beta^2}{\partial x^\mu} &=& \left(\frac{\partial b_a}{\partial x^\mu}b_b+b_a\frac{\partial b_b}{\partial x^\mu}\right)y^a y^b,\\
		\frac{\partial \alpha\beta}{\partial x^\mu} &=& b_c y^c\frac{\partial \alpha}{\partial x^\mu}+ \alpha\frac{\partial b_c}{\partial x^\mu}y^c.
	\end{eqnarray}
\end{subequations}
Using the definition of the geodesic coefficient in (\ref{FG3}) and above expressions (\ref{DDF1}) and (\ref{DDF2}), we can obtain
\begin{equation}\label{DDL1}
	\frac{\partial\mathcal{L}^2}{\partial x^\nu}=
	m^2\frac{\partial \widetilde{g}_{ab}}{\partial x^\nu}y^a y^b
	+y^a y^b\left(\frac{\partial b_a}{\partial x^\nu}b_b+b_a\frac{\partial b_b}{\partial x^\nu}\right)
	+2b_c y^c \frac{\partial \alpha}{\partial x^\nu}+2\alpha\frac{\partial b_c}{\partial x^\nu}y^c
\end{equation}
and
\begin{eqnarray}\label{DDL2}
	\frac{\partial^2 \mathcal{L}^2}{\partial y^k\partial x^\nu}&=&
	m^2\left(\frac{\partial \widetilde{g}_{kb}}{\partial x^\nu}y^b+\frac{\partial \widetilde{g}_{ak}}{\partial x^\nu}y^a\right)
	+2b_b\left(\frac{\partial b_k}{\partial x^\nu}y^b+y^k\frac{\partial b_b}{\partial x^\nu}\right) \nonumber\\
	&+& 2b_k \frac{\partial \alpha}{\partial x^\nu}+2b_c y^c \frac{\partial^2 \alpha}{\partial y^k\partial x^\nu}
	+2y^c \frac{\partial \alpha}{\partial y^k}\frac{\partial b_c}{\partial x^\nu}+2\alpha\frac{\partial b_k}{\partial x^\nu}.
\end{eqnarray}
By combining (\ref{DDL1}) and (\ref{DDL2}) and rearranging the similar terms, the geodesic coefficient can be obtained\cite{Zhou}
\begin{equation}\label{EMTb}
	G^\mu(x,v)=\frac{1}{4}g^{\mu\nu}\left(-2e\alpha F_{\nu\sigma}+Q_{\nu\lambda\sigma}y^\lambda
	+M_{\nu\lambda\sigma}y^\lambda+S_{\nu\lambda\sigma}y^\lambda\right)y^\sigma,
\end{equation}
where
\begin{subequations}\label{EMF3}
	\begin{eqnarray}
		F_{\nu\sigma}&=& \frac{\partial A_\sigma}{\partial x^\nu}-\frac{\partial A_\nu}{\partial x^\sigma},\\
		Q_{\nu\lambda\sigma}&=& m^2\left(\frac{\partial \widetilde{g}_{\nu \lambda}}{\partial x^\sigma}+\frac{\partial \widetilde{g}_{\lambda\nu}}{\partial x^\sigma}
		-\frac{\partial \widetilde{g}_{\lambda\sigma}}{\partial x^\nu} \right),\\
		M_{\nu\lambda\sigma}&=& \frac{em}{\widetilde{\alpha}}\left(A_\nu\frac{\partial}{\partial x^\sigma}-A_\sigma\frac{\partial}{\partial x^\nu}\right)
		\widetilde{g}_{\lambda\sigma}y^\sigma, \\
		S_{\nu\lambda\sigma}&=& 2e^2\frac{\partial \mathcal{A}_{\nu\sigma}}{\partial x^\lambda}-e^2\frac{\partial \mathcal{A}_{\lambda\sigma}}{\partial x^\nu}
		+2e\frac{\partial B_{\lambda\nu}}{\partial x^\sigma},
	\end{eqnarray}
\end{subequations}
where $F_{\nu\sigma}$ is the electromagnetic field.
$Q_{\nu\lambda\sigma}$ depends on mass and curved spacetime.
$M_{\nu\lambda\sigma}$ describes a coupled effect between the curved spacetime and electromagnetic field.
$S_{\nu\lambda\sigma}$ is a pure effect emerged in the Finsler-Randers space, in which
$\mathcal{A}_{\nu\sigma}=A_\nu A_\sigma$ describes a coupling of electric and magnetic fields and
$B_{\lambda\nu}=A_\lambda \ell_\nu$ is a coupling of the electromagnetic field and velocity field.
In the flat spacetime,
$Q_{\nu\lambda\sigma}=M_{\nu\lambda\sigma}=0$. When $A_\mu$ vanishes, $F_{\nu\sigma}=M_{\nu\lambda\sigma}=S_{\nu\lambda\sigma}=0$,
but $Q_{\nu\lambda\sigma}\neq 0$.

\subsubsection{Berwald space}
In fact, the geodesic coefficient can be rewritten as another expression for giving the Berwald structure. \cite{Bao}
\begin{equation}\label{GGa}
	G^\lambda=\frac{1}{2}\gamma^\lambda\ _{\mu\nu} y^\mu y^\nu,
\end{equation}
where
\begin{equation}\label{GGc}
	\gamma^i\ _{jk} = \frac{1}{2}g^{i \ell}\left(\frac{\partial g_{\ell k}}{\partial x^j}
	+\frac{\partial g_{j\ell}}{\partial x^k}-\frac{\partial g_{jk}}{\partial x^\ell}\right).
\end{equation}

One introduces \cite{Bao}
\begin{equation}\label{Bb0}
	B_{\mu|\nu}:=\frac{\partial b_\mu}{\partial x^\nu}-b_\sigma\widetilde{\gamma}^\sigma\ _{\mu\nu},
\end{equation}
where
\begin{equation}\label{GGc}
	\widetilde{\gamma}^\lambda\ _{\mu\nu}=\frac{1}{2}\widetilde{g}^{\lambda \sigma}\left(\frac{\partial \widetilde{g}_{\sigma\nu}}{\partial x^\mu }
	+\frac{\partial \widetilde{g}_{\mu\sigma}}{\partial x^\nu}-\frac{\partial \widetilde{g}_{\mu\nu}}{\partial x^\sigma}\right).
\end{equation}
By a straight forward computation, one gets\cite{Bao}
\begin{eqnarray}\label{GGd}
	\frac{\mathcal{L}^2}{\alpha^2}\gamma^i\ _{jk} \ell^j \ell^k &=& \widetilde{\gamma}^i\ _{jk}\ell^j \ell^k
	+m^2B_{j|k}\left(\widetilde{g}^{ij}\widetilde{\ell}^k-\widetilde{g}^{ik}\widetilde{\ell}^j \right)   \nonumber\\
	&+& B_{j|k}\ell^i \left(\widetilde{\ell}^j\widetilde{\ell}^k+\widetilde{\ell}^j\widetilde{b}^k-\widetilde{\ell}^k \widetilde{b}^j\right),
\end{eqnarray}
where $\ell^i=\frac{y^i}{\mathcal{L}}$, $\widetilde{\ell}^i=\frac{y^i}{\alpha}$, and $\widetilde{b}^i=\frac{1}{m^2}\widetilde{g}^{ij}b_j$.
Thus, the geodesic coefficient can be expressed as two equivalent forms,\cite{Bao,Zhou}
\begin{eqnarray}\label{GGe}
	2G^i &=& \gamma^i\ _{jk} y^j y^k \\
	&=&\widetilde{\gamma}^i\ _{jk} y^j y^k++m^2B_{j|k}\left(\widetilde{g}^{ij}y^k-\widetilde{g}^{ik}y^j \right)\alpha   \nonumber\\
	&+& B_{j|k}\ell^i \left[y^jy^k+\left(y^j\widetilde{b}^k-y^k \widetilde{b}^j\right)\alpha\right].
\end{eqnarray}
It can be seen that when $B_{j|k}=0$, we have
\begin{equation}\label{GGf}
	G^i = \frac{1}{2}\gamma^i\ _{jk} y^j y^k= \frac{1}{2}\widetilde{\gamma}^i\ _{jk} y^j y^k.
\end{equation}
The Finsler-Randers structure with $B_{j|k}=0$ is so-called the Berwald structure.
From the physical viewpoint, the Berwald condition can be regarded as a constraint on the subspace in the Finsler-Randers space, where the $hv$ part of the Chern curvature vanishes.
\begin{equation}\label{Bb2}
	\frac{\partial A_\mu}{\partial x^\nu}-A_\sigma\widetilde{\gamma}^\sigma\ _{\mu\nu}=0.
\end{equation}
This gauge condition (\ref{Bb2}) ensure that the geodesic coefficient has no $v$-dependence. This property
is referred to as the Randers space with the Berwald structure.\cite{Bao} Physically, this gauge can be interpreted as a weak field approximation.

\subsubsection{Derivation of generalized Maxwell equations}
The Lagrangian in the Finsler-Randers space can be given by
\begin{equation}\label{LL1}
		\mathcal{L} = -\frac{\sqrt{g}}{4}\mathcal{F}_{\mu\nu}\mathcal{F}^{\mu\nu}+\sqrt{g}A_\mu J^\mu 
\end{equation}
where $\mathcal{F}_{\mu\nu}=F_{\mu\nu}+F^G_{\mu\nu}$ and $g=\det{g_{\mu\nu}}$.
The Euler-Lagrange equation is given by
\begin{equation}\label{Lqe0}
	\nabla_\nu\frac{\partial \mathcal{L}}{\partial (\partial_\nu A_\mu)}=\frac{\partial \mathcal{L}}{\partial A_\mu},
\end{equation}
where $\nabla_\mu$ is the covariant derivative in the horizontal direction, which is defined by 
\begin{equation}
	\nabla_\mu \mathcal{F}^{\mu\nu}=\frac{\delta \mathcal{F}^{\mu\nu}}{\delta x^\mu}
	+\mathcal{F}^{\kappa\nu}\Gamma^\mu\ _{\kappa\mu}
	+\mathcal{F}^{\mu\kappa}\Gamma^\nu\ _{\kappa\mu}
\end{equation}
where $\Gamma^\mu\ _{\kappa\mu}$ is the Chern connection coefficient and
\begin{equation}
	\frac{\delta }{\delta x^\mu}=\frac{\partial}{\partial x^\mu}-N^\kappa\ _\mu \frac{\partial}{\partial y^\kappa} 
\end{equation}
where $N^\kappa\ _\mu$ is the Cartan connection coefficient. 
Using the basic formula,
\begin{subequations}\label{LLLA1}
	\begin{eqnarray}
		\frac{\partial \mathcal{L}}{\partial A_\mu} &=& J^\mu, \\
		\frac{\partial F_{\alpha\beta}}{\partial(\partial_\nu A_\mu)} &=& 
		\delta_{\alpha\nu}\delta_{\beta\mu}-\delta_{\alpha\mu}\delta^{\beta\nu}, \\	
		\frac{\partial F^{\alpha\beta}}{\partial (\partial_\nu A_\mu)} &=&
		g^{\alpha\nu}g^{\beta\mu}-g^{\alpha\mu}g^{\beta\nu}, \\	
		\frac{\partial F_{\alpha\beta}F^{\alpha\beta}}{\partial (\partial_\nu A_\mu)} 
		&=& F^{\nu\mu}, \\			
		\frac{\partial \mathcal{F}_{\alpha\beta}\mathcal{F}^{\alpha\beta}}{\partial (\partial_\nu A_\mu)} &=& 2\mathcal{F}^{\left[\nu\mu\right]}
		+\frac{\partial \mathcal{F}^G_{\alpha\beta}}{\partial (\partial_\nu A_\mu)}
		\mathcal{F}^{\alpha\beta}+\mathcal{F}_{\alpha\beta}\frac{\partial \mathcal{F}^{G\alpha\beta}}{\partial (\partial_\nu A_\mu)}, 
	\end{eqnarray}
\end{subequations}
and substituting them into the Euler-Lagrange equation, we get the Maxwell equation (\ref{GMWEq1}).

\begin{equation}
	\nabla_\mu F^{\mu\nu}+\nabla_\mu \mathcal{F}^{G\left[\mu\nu\right]}
	-\frac{1}{4}\nabla_\mu\left(\frac{\partial F^G_{\alpha\beta}}{\partial(\partial_\nu A_\mu)} \mathcal{F}^{\alpha\beta}+\mathcal{F}_{\alpha\beta}\frac{\partial F^{G\alpha\beta}}{\partial (\partial_\nu A_\mu)}\right) =J^\nu.
\end{equation}
The Maxwell equations involve the complicated Finslerian structure, 
to get the basic physics of the geometric field, let us ignore the detail of the Finslerian structure, namely, $\nabla_\nu\approx \frac{\partial}{\partial x^\nu}$, we have
\begin{eqnarray}
	\frac{\partial}{\partial x_\nu}\left(\frac{\partial F^G_{\alpha\beta}}{\partial(\partial_\nu A_\mu)} \mathcal{F}^{\alpha\beta}+\mathcal{F}_{\alpha\beta}\frac{\partial F^{G\alpha\beta}}{\partial (\partial_\nu A_\mu)}\right) 
	&=& 
	2\mathcal{F}^{\rho\sigma}
	\frac{\partial^2 F^G_{\rho\sigma}}{\partial x^\nu(\partial_\nu A_\mu)} \nonumber \\
	&-&\left( \frac{\partial \mathcal{F}^{\rho\sigma}}{\partial x^\nu}
	\frac{\partial F^G_{\rho\sigma}}{\partial (\partial_\nu A_\mu)}
	+\frac{\partial \mathcal{F}_{\rho\sigma}}{\partial x^\nu}
	\frac{\partial F^{G\rho\sigma}}{\partial (\partial_\nu A_\mu)}\right) 
\end{eqnarray}
Consequently, we get (\ref{EMeq2}b).
	
	% =========================
	% 参考文献
	% =========================
%	\bibliographystyle{unsrt}
%	\bibliography{references}
\bibliography{apssamp}
% Produces the bibliography via BibTeX.

\end{document}